\documentclass{article}
\usepackage{natbib}
\setcitestyle{numbers,square}
\usepackage{neurips_2024}
\usepackage{graphicx}
\usepackage{wrapfig}
\usepackage{multirow}
\usepackage{caption}
\usepackage{subcaption}
\usepackage{array}

\usepackage[utf8]{inputenc} 
\usepackage[T1]{fontenc}    
\usepackage{hyperref}       
\usepackage{url}            
\usepackage{booktabs}       
\usepackage{amsfonts}       
\usepackage{nicefrac}       
\usepackage{microtype}      
\usepackage{xcolor}         

\title{Harmonizing Pixels and Melodies: Maestro-Guided Film Score Generation and Composition Style Transfer}

\author{%
  \thanks{Use footnote for providing further information
    about author (webpage, alternative address)---\emph{not} for acknowledging
    funding agencies.} \\
  Department of Computer Science\\
  Cranberry-Lemon University\\
  Pittsburgh, PA 15213 \\
  \texttt{hippo@cs.cranberry-lemon.edu} \\
}

\begin{document}
\maketitle

\begin{abstract}
We introduce a film score generation framework to harmonize visual pixels and music melodies utilizing a latent diffusion model. Our framework processes film clips as input and generates music that aligns with a general theme while offering the capability to tailor outputs to a specific composition style. Our model directly produces music from video, utilizing a streamlined and efficient tuning mechanism on ControlNet. It also integrates a film encoder adept at understanding the film's semantic depth, emotional impact, and aesthetic appeal. Additionally, we introduce a novel, effective yet straightforward evaluation metric to evaluate the originality and recognizability of music within film scores.
To fill this gap for film scores, we curate a comprehensive dataset of film videos and legendary original scores, injecting domain-specific knowledge into our data-driven generation model. Our model outperforms existing methodologies in creating film scores, capable of generating music that reflects the guidance of a maestro's style, thereby redefining the benchmark for automated film scores and laying a robust groundwork for future research in this domain. The code and generated samples are available at \href{https://anonymous.4open.science/r/HPM-C7DC}{https://anonymous.4open.science/r/HPM}.
\end{abstract}

\section{Introduction}
A film score - the original music accompanying a film \/- plays a pivotal role in enriching the film's emotional landscape, deepening narrative complexity, character arcs, and thematic exploration. 
Creating a film score requires a harmonious, multidisciplinary effort that includes composers, orchestrators, musicians, sound engineers, and music editors working in concert to blend composition, arrangement, and recording with the film's visual narrative.  
Automating the film score production process through artificial intelligence research represents a significant stride toward cost efficiency and innovation in film score production.

Translating the visual modality into music has been a promising area in cross-modal generative modeling~\cite{zhao2014emotion,di2021video,zhu2022discrete,Yu2023Long}. Some works rely on Codebook~\cite{copet2023simple} and have less flexibility to generate novel sounds outside it.
Some works use pre-defined symbolic musical representations such as MIDI (Musical Instrument Digital Interface), REMI (revamped MIDI-derived events), and Piano-Roll that can be autoregressively generated~\cite{dong2018musegan,di2021video,gan2020foley}. 
Nevertheless, they require an excellent MIDI synth to render audio from generated MIDI and struggle to model timbre and expressiveness. Recently, some works directly produce spectrograms based on diffusion models~\cite{Forsgren_Martiros_2022,ghosal2023text,liu2023audioldm} with textual prompts.
Diffusion model-generated audio exhibits fidelity to prominent prompt features, including genre, tempo, instrumentation, and mood, while also capturing the fine-grained semantics of the prompt.
Yu et al.~\cite{Yu2023Long} utilize a latent conditional diffusion probabilistic model to synthesize long-term conditional waveforms and generate soundtracks for dances and sports scenarios.
Recent work\footnote{https://huggingface.co/spaces/fffiloni/image-to-music-v2} involves translating visual content into textual descriptions~\cite{li2023blip,li2023llama}, utilizing robust text-to-music diffusion models~\cite{huang2023noise2music,liu2023audioldm,ghosal2023text} for music generation. However, this two-step process (visual-to-text and then text-to-music) introduces additional complexity and points of potential error. Key visual elements crucial for emotion conveyance, like colors, lighting, and composition, may be inadequately represented in text. In contrast, a direct visual-to-music translation could more effectively capture and convert these visual cues into musical elements, preserving the original emotional tone of the visuals.

While conceptually straightforward, generating music from film diffusion models faces notable challenges. 
1) The field significantly lacks datasets that carefully pair film clips with their corresponding music. Compiling such datasets is challenging and resource-intensive.
2) Achieving the thematic musical pieces align with the film's narrative and emotional tone presents a complex challenge, introducing integration difficulties within the current frameworks of diffusion models.
3) There is an absence of objective metrics to measure the quality of music generated for film clips, complicating the evaluation of progress and the refinement of models.

To bridge the existing void in automated film scores, we establish a comprehensive dataset - FilmScoreDB. FilmScoreDB contains 32,520 film clip-music pairs, totaling 90.35 hours, featuring compositions from renowned film composers. This collection serves to infuse our data-driven diffusion model with targeted domain-specific insights. We present HPM, a novel approach tailored for generating film scores and transferring composition styles.
Leveraging a diffusion model, our framework introduces a low-rank, parameter-efficient fine-tuning mechanism to ControlNet, complemented by a film encoder designed to assimilate semantic, emotional, and aesthetic dimensions of film content.
Additionally, we introduce an enhanced metric incorporating originality and recognizability to assess the quality of generated music.
Utilizing the design above, our model framework adeptly generates film scores, operating independently and under specific styles' control, demonstrating a significant advancement in film score generation.
In conclusion, our main contributions are three-fold:
\begin{itemize}
    \item 
We are the first to tackle the automatic film score and composition style transfer challenge, concentrating on generating music for specific film segments.
    \item We introduce a novel benchmark that includes a comprehensive film score dataset, a refined set of evaluation metrics, and a well-defined baseline model to foster subsequent research efforts.
    \item Through comprehensive experimental analysis, our framework demonstrates exceptional capability in generating film scores and in transferring composition styles.
    Our framework outperforms existing methods across all evaluated metrics and sets a new benchmark for the musical arts community.
\end{itemize}

\section{Dataset}
\label{sec:dataset}
As shown in Tab.~\ref{tab-dataset}, diverse video-music datasets serve specialized research needs. Datasets focusing on dance~\cite{li2021ai} and skating~\cite{xu2019learning} explore the alignment of music rhythm with body motion. \begin{wraptable}{r}{7cm}
 \caption{Diverse Video-Music Datasets.}
\label{dataset}
\label{tab-dataset}
\centering
    \scalebox{0.55}{\begin{tabular}{@{}llllll@{}}
    \toprule
\textbf{Dataset} & \textbf{Video Content}  & \textbf{Size}  & \textbf{Open} & \textbf{Video} & \textbf{Total Duration} \\ \midrule
URMP~\cite{li2018creating} & Percussion Performance  & 44  & \checkmark  & \checkmark &1.3h \\
AtinPiano~\cite{moryossef2019your}  & Piano Performance  & 257 & \checkmark & \checkmark & 17.2h        \\
MUSIC \cite{zhao2018sound} & Performance Video  & 685  & \checkmark  & \checkmark & 45.7h \\
EmoMV ~\cite{thao2023emomv} & Music Video   & 5,986 & \checkmark & \checkmark & 44.33h \\
MuVi ~\cite{chua2022predicting}& Music Video  & 811 & \checkmark  & \checkmark & 13.52h \\
SymMV ~\cite{zhuo2023video}  & Music Video  & 1,140  & \checkmark  & $\times$ & 76h\\
MVDB ~\cite{pandeya2021deep}& Music Video  & 1,985 & \checkmark & \checkmark & 16.54h \\
FS1000 ~\cite{xie2022vector}  & Skate Video  & 1,604 & \checkmark& \checkmark & 89.1h\\
FisV~\cite{xu2019learning} & Skate Video  & 500 & \checkmark  & \checkmark  & 23.6h \\
AIST++ ~\cite{lin2021exploring} & Dance Video  & 60   & \checkmark  & \checkmark  &4h         \\
TikTok ~\cite{lin2014microsoft}  & Dance Video  & 445  & \checkmark & \checkmark &1.55h \\ \midrule
\textbf{FilmScoreDB} & \textbf{Film Video}  & \textbf{32,520} & \textbf{\checkmark}& \textbf{\checkmark} & \textbf{90.35h}  \\\bottomrule
\end{tabular}}
\end{wraptable} Conversely, music performance datasets~\cite{zhao2018sound} aim to synchronize the human body and finger movements with multi-instrumental music. Using music video datasets for video-based music generation aligns with our goal of thematic and emotional resonance. 
However, music video datasets differ significantly from film score datasets,
with fundamental differences in objectives: Music video directors visually promote songs, whereas film score composers enhance a film's emotional and narrative depth. 
To further elaborate, there are some specific differences.
\textbf{Technical Complexity}: Film scores require deep musical theory, orchestration, and narrative alignment, contrasting with music videos' emphasis on visual creativity and editing skills.
\textbf{Cultural and Historical Significance}: Film scores, drawing from classical music traditions, possess significant historical depth and cultural significance, in contrast to music videos that are shaped by and contribute to contemporary trends. 
Therefore, developing large-scale, high-quality datasets characterized by renowned film composers can drive research on video music.

For this purpose,  we compile FilmScoreDB, a comprehensive dataset featuring film video clips and their corresponding legendary original scores. 
The collected FilmScoreDB contains 32,520 samples, sourced from nearly 300 famous films worldwide, each 10 seconds long. We split FilmScoreDB into a training set (26,730 pairs), a validation set (2,895 pairs), and a test set (2,895 pairs). We start by gathering a list of films with Best Original Score Nominees. To obtain such paired data, we correctly obtain and download about 300 films on platforms such as YouTube, Netflix, Disney+, Prime Video, etc., invited five connoisseurs to watch, and extract film clips, including scores. In addition, each data pair is labeled with relevant film score composers and styles. For this dataset, we only provide relevant information about the films used, including film names, composers, score information, and clip timestamps. We also evaluate our model using the EmoMV~\cite{thao2023emomv} dataset to compare it fairly with prior work.
\begin{figure}[tb]
  \centering
  \includegraphics[height=6.5cm]{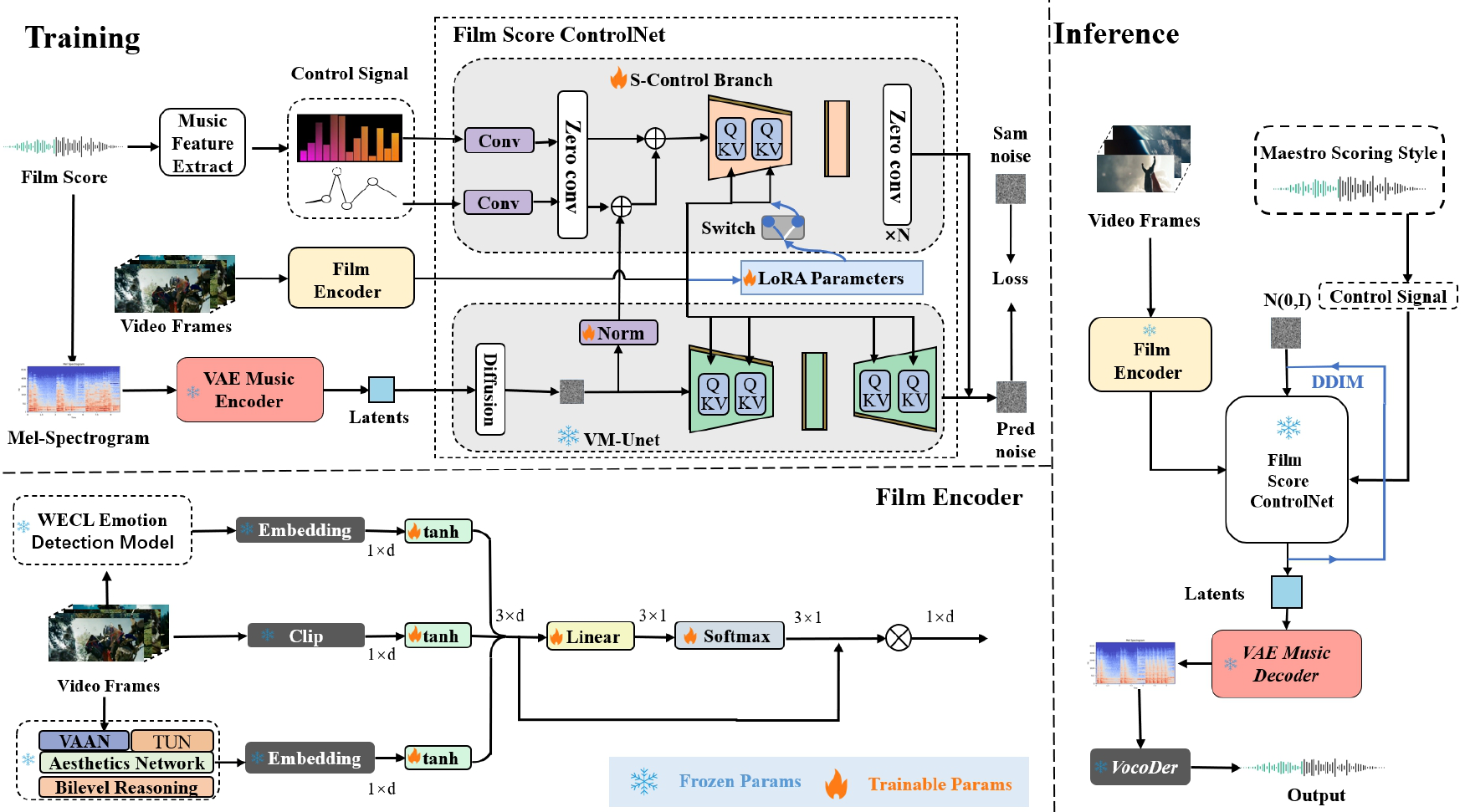}
  \caption{Illustration of our HPM framework. a) During the training stage, our model incorporates the video feature as a global control input, alongside the local control signal of melody and dynamics. b) During the inference stage, the local controls can be one composition style of one specific master, guiding the Film Score ControlNet to produce a Mel-spectrogram, subsequently converted into audio via a vocoder. c) The film encoder processes to extract emotional, semantic, and aesthetic embeddings, enriching the model's interpretative depth.}
  \label{fig:framework}
\end{figure}

\section{Methodology}
Our overall goal is to learn a conditional generative model, $p(s|c_{film}, C_{style})$, for film scores. Here, $c_{film}$ represents the conditioning information obtained from a given film clip through the Film Encoder (time-independent), and $C_{style}$ is a set of music style control signals (melody, dynamic). We incorporate the obtained music style control signals into a ControlNet, inspired by the Uni-ControlNet~\cite{zhao2024uni}, and optimize the model training using the low-rank adaptation (LORA)~\cite{hu2021lora}. We generate new film scores with different styles based on the conditioning $c_{film}$, and the music style control signals $C_{style}$ during inference.

\subsection{Film Encoder}
We extract semantic, aesthetic, and emotional features from film videos as model guidance to mitigate the complexity and computational demands of using raw video frames for music generation. 
Semantic features are derived using the image encoder of CLIP~\cite{radford2021learning}, averaging extracted frame features to capture video semantics.
Aesthetic features, assessing visual qualities like lighting and color, are obtained using a pre-trained TAVAR model~\cite{li2023theme}, which evaluates visual attributes and theme-related aesthetics through a combination of networks and feature mapping. 
For emotional features, we employ the pre-trained WECL~\cite{zhang2023weakly} model with the subjective nature of emotions in videos by utilizing intra-modal and inter-modal relationships.
We employ a lightweight attention feature fusion method to integrate these diverse features effectively, optimizing music generation by learning fusion weights for a harmonized feature combination.
Due to space constraints, the detailed process for this section is included in the supplementary materials.

\subsection{Film Score ControlNet}
To realize film score generation and Maestro-guided Composition Style Transfer,
inspired by Uni-ControlNet~\cite{zhao2024uni}, we propose Film Score ControlNet, a unified framework that allows for the simultaneous utilization of different local controls (e.g., melody and dynamic) and global controls (e.g., video semantic features, emotional features, and aesthetic features) in a single model, enabling flexible and compositional style-controllable film score generation.

In this task, given the global video control $c_{film}$ and a set of local controls $C_{style}$ that control the music style, our overall objective is to learn a conditional generative model $p_{\theta}(z_0|c_{film}, C_{style})$ on the compressed representation $z$ of the music mel-spectrogram $X$.
\subsubsection{Foundational Film Score Generation Model}
We utilize a probabilistic generative model to estimate the true conditional data distribution $q(z_0|c_{film})$, where the model distribution is denoted as $p_{\theta}(z_0|c_{film})$. Here, $z_0 \in R^{C\times\frac{T}{r}\times\frac{F}{r}}$ represents the prior audio sample $x$ in the latent space compressed from the mel-spectrogram $X \in R^{T\times F}$, where $r$ denotes the compression level, $C$ represents the channels of the latent representation, $T$ and $F$ represent the time and frequency dimensions of the mel-spectrogram $X$, respectively.

The diffusion model consists of two processes: a forward process that gradually adds a small amount of high-level Gaussian noise to the sample $z_0$ over $M$ steps, and a corresponding backward process that predicts the added noise during the forward process and eliminates the noise to recover the input latent.
To enhance the synchronization between video and music, we fine-tune the AudioLDM~\cite{liu2023audioldm}  employing a dedicated video-to-music dataset, HIMV-200k~\cite{hong2017content}.
This optimization transitions the foundation model from a text-music space to a video-music space. We refer to this pre-trained model as \textbf{VM-Unet} in the following.

\subsubsection{Adding Style Controls to Foundational Model}
As shown in Fig.~\ref{fig:framework}, similar to ControlNet~\cite{zhao2024uni}, we fix the weights of \textbf{VM-Unet}. We replicate the structure and weights of the Down Block and Middle Block from \textbf{VM-Unet} and add some new zero-convolution layers, called \textbf{S-Control Branch}.
During the training stage, we take the music style of melody and dynamic as the local control, and film feature as global control into S-Control Branch.
The compressed representation of the mel-spectrogram is fed into \textbf{VM-Unet}. Only the parameters of the \textbf{S-Control Branch} are trainable. 

Specifically, let $f(x^{(m,l-1)},m,c_{film},C_{style})$ denote the $l^{th}$ block of the \textbf{S-Control Branch}, where $m$ is the diffusion time step, $x^{(m,l-1)}$ contains the features of the noised latent after $l-1$ blocks, and $c_{film}$, $C_{style}$ are the global and  local control, respectively. The style control is incorporated via:
\begin{equation}
    \check{f}^{(l)}(x^{m,l-1},m,c_{film},C_{style}) := Z_{out}(f^l(x^{(m,l-1)}+C_{style},m,c_{film})),
\end{equation}
where $Z_{out}$  is the newly attached zero convolution
layer and $f^{(l)}$ is initialized from the $l^{th}$ encoder block of the
pre-trained video-conditioned UNet.

For global control, cross-attention layers are employed to capture visual information $c_{film}$ from the input film clip. In our setup, $z_t$ represents incoming noise features, $W_q$, $W_k$, and $W_v$ are projection matrices. The $Q$, $K$, and $V$ in cross-attention can be denoted as:
\begin{equation}
Q = W_q(z_t), K = W_k(c_{film}), V = W_v(c_{film}).
\end{equation}
 
For local control, we start by aligning the melody and dynamic control signals with the input noise features' resolution using two convolutional blocks and a newly added zero convolution layer. Then, we integrate them with the normalized input noise features through a shortcut in the following sequence:
\begin{equation}
C = norm(z_t)+Z_{in}(conv_{mel}(c_{mel}))+Z_{in}(conv_{dyn}(c_{dyn})),
\end{equation}
where $Z_{in}$ is also a new input zero convolution layer.
For Melody$(c_{mel} \in R^{T\times12})$, we adapt the chromagram using a window size of 260 and a hop size of 160, condensing energy from $F$ frequency bins into 12 pitch classes.
Then, we enhance it by selecting the most dominant pitch class through argmax at each time step. For Dynamic$(c_{dyn} \in R^{T\times1})$, we derive loudness by summing energy from frequency bins in each time frame of the linear spectrogram and converting the values to decibels (dB), which aligns closely with human perception of loudness. To reduce rapid fluctuations caused by note onsets or percussion, and to align our dynamic control with perceived musical intensity, we smooth the values by using a second-order context window, namely the Savitzky-Golay filter~\cite{virtanen2020fundamental}.

\subsubsection{Low-Rank Adaptation For Improving Training}               
Despite ControlNet being half the size of the stable diffusion model, it still possesses too many parameters for consumer GPUs to handle efficiently.
The LoRA~\cite{hu2021lora} is a low-resource fine-tuning approach for large models.
Given a pre-trained model layer with weights $W_0 \in R^{d\times k}$, where $d$ is the input dimension and $k$ is the output dimension, LoRA decomposes $\Delta W$ as:
\begin{equation}
    \Delta W = BA,
\end{equation}
where $B \in R^{d\times r}$ and $A \in R^{r\times k}$ with $r \ll min(d, k)$. $min(d, k)$ is a small rank that constrains the update to a low dimensional
subspace. By freezing $W_0$ and only
optimizing the smaller
matrices $A$ and $B$, LoRA achieves massive reductions in
trainable parameters. During inference, $\Delta W$ can be merged
into $W_0$ with no overhead by a LoRA scaling factor $\alpha$:
\begin{equation}
W = W_0 + \alpha\Delta W.
\end{equation}
Subsequently, we inject trainable layers (low-rank decomposition matrices) into each transformer block in the \textbf{S-Control Branch}.  Experimental results demonstrate that utilizing LoRA to fine-tune enhances the efficiency of the process, resulting in a faster performance with a reduction in computational demand.

\subsubsection{Inference}  
During inference, the condition of local style control is set to zero for film score generation. 
For composition style transfer, we can control the style based on melody, dynamics, or both, allowing for flexible adaptation of the composition's style to the desired characteristics.
\section{Experiments}
This section provides a detailed evaluation of our Harmonized Pixel and Melody Film Score Diffusion (HPM) model, integrating pixel-level and melody-level aspects. We employ objective and subjective metrics on two different datasets to assess our model. To evaluate generated music's originality, we introduce the \textit{originality vs. recognizability} framework~\cite{boutin2022diversity}. Additionally, we demonstrate the effectiveness of using LoRA for training acceleration within our framework.

\subsection{Implementation Details}
Considering the computational efficiency, we use AudioLDM~\cite{liu2023audioldm} as our basic backbone. We fix the parameters of Music VAE in the AudioLDM.
For all visual features, we use a frame rate of 10 fps, following the standard on Film Score (ours) and EmoMV. We use the default hyper-parameters of AudioLDM and employ a base learning rate of $0.0001$.  The film score generation diffusion model training over $200,000$ gradient steps with a batch size of $2$ is executed on $4$ NVIDIA GeForce RTX $3090$ GPUs, taking approximately $40$ hours to complete over $60$ epochs. We apply the AdamW optimizer~\cite{loshchilov2017decoupled} with $\beta_1 = 0.9$, $\beta_2 = 0.999$, and a weight decay of $0.01$. During inference stage, we adopt DDIM~\cite{song2020denoising}  for sampling, with the number of timesteps set to $200$ and the classifier free guidance scale~\cite{ho2022classifier} set to $7.5  $. 
We use the L2 loss and AdamW optimizer for fine-tuning our Film Score ControlNet until convergence for $12$ hours, using $4$ NVIDIA GeForce RTX $3090$ GPUs. During inference stage, we use 100-step DDIM sampling and only use classifier-free guidance on global video control. Due to the constraints of paper length, details regarding the film encoder are provided in the supplementary materials.

\subsection{Evaluation Methods}

\subsubsection{Baseline}
For a comprehensive evaluation of our HPM framework, we conduct comparative experiments on two tasks.
For \textbf{film score generation}, we select seven high-performing methods with available code as baselines. 
Specifically, we compare our HPM with LORIS~\cite{Yu2023Long}, D2M-Gan model~\cite{zhu2022quantized}, CDCD model~\cite{zhu2022discrete} based on Codebook, and DIFF-Foley model~\cite{luo2024diff} based on Mel-Spectrogram, we employ the official implementation. 
We further fine-tune a video-to-music generation model, dubbed Tango-VM, building upon the text-to-music model, Tango~\cite{ghosal2023text}.
Due to the difficulty of converting film music into MIDI without annotations, we exclude MIDI-generation-based methods, such 
as CMT~\cite{di2021video}, Foley Music~\cite{gan2020foley}, Audeo~\cite{su2020audeo}, and Video2Music~\cite{kang2023video2music} from our baseline comparisons.
For \textbf{Composition Style Transfer}, we compare our model with the DITTO~\cite{novack2024ditto} trained with melody and dynamics controls.
In EmoMV, each emotion is designated as a distinct style. Similarly, each film score style within FilmScoreDB is identified as a unique style.
The DITTO model realizes text-driven music stylization, leveraging melody and intensity conditioning through inference-time optimization of the initial noise latent space in a pre-trained text-to-music diffusion model.  Inspired by DITTO, we evaluate our VM-net model across three scenarios: (i) melody-only, excluding melodies, (ii) dynamics-only, excluding dynamics, and (iii) combined melody and dynamic control.
\subsubsection{Evalation Metric}
\textbf{Objective}: To assess rhythm correspondence in our study, we utilize modified versions of the Beats Coverage Score (BCS)~\cite{davis2018visual}, Beats Hit Score (BHS)~\cite{lee2019dancing}, and F1 scores for a thorough evaluation of rhythm accuracy, CSD and HSD (the standard deviations of BCS and BHS) to evaluate generation stability.
For music quality, the Frechet Audio Distance (FAD)~\cite{kilgour2019frechet}, Inception Score, and Kullback-Leibler (KL)~\cite{hershey2017cnn} divergence are employed, leveraging the VGGish classifier for FAD to measure similarity and PANN for IS and KL to evaluate quality.
For style transfer, we measure Melody Accuracy and Dynamics Correlation~\cite{wu2023music}, analyzing the alignment of pitch classes and the Pearson correlation of dynamics between input controls and generated output, with micro and macro correlations providing insights into dynamic control fidelity and consistency across generations.

\textbf{Subjective}: In addition, we also used the subjective measure of Mean Opinion Score (MOS) in the video-to-music generation task. MOS is obtained by calculating the average score of user research on overall music quality. 
Here, we recruit 20 participants to listen to 30 pieces of music generated by HPM and five baseline methods (5 samples per method) and then evaluate the overall music quality by giving scores ranging from 1 to 5 in terms of music quality, clarity, and presence or absence of delay.

\subsubsection{The Originality vs. Recognizability Framework}
\begin{wrapfigure}{r}{8cm}
\centering
  \includegraphics[height=3.7cm]{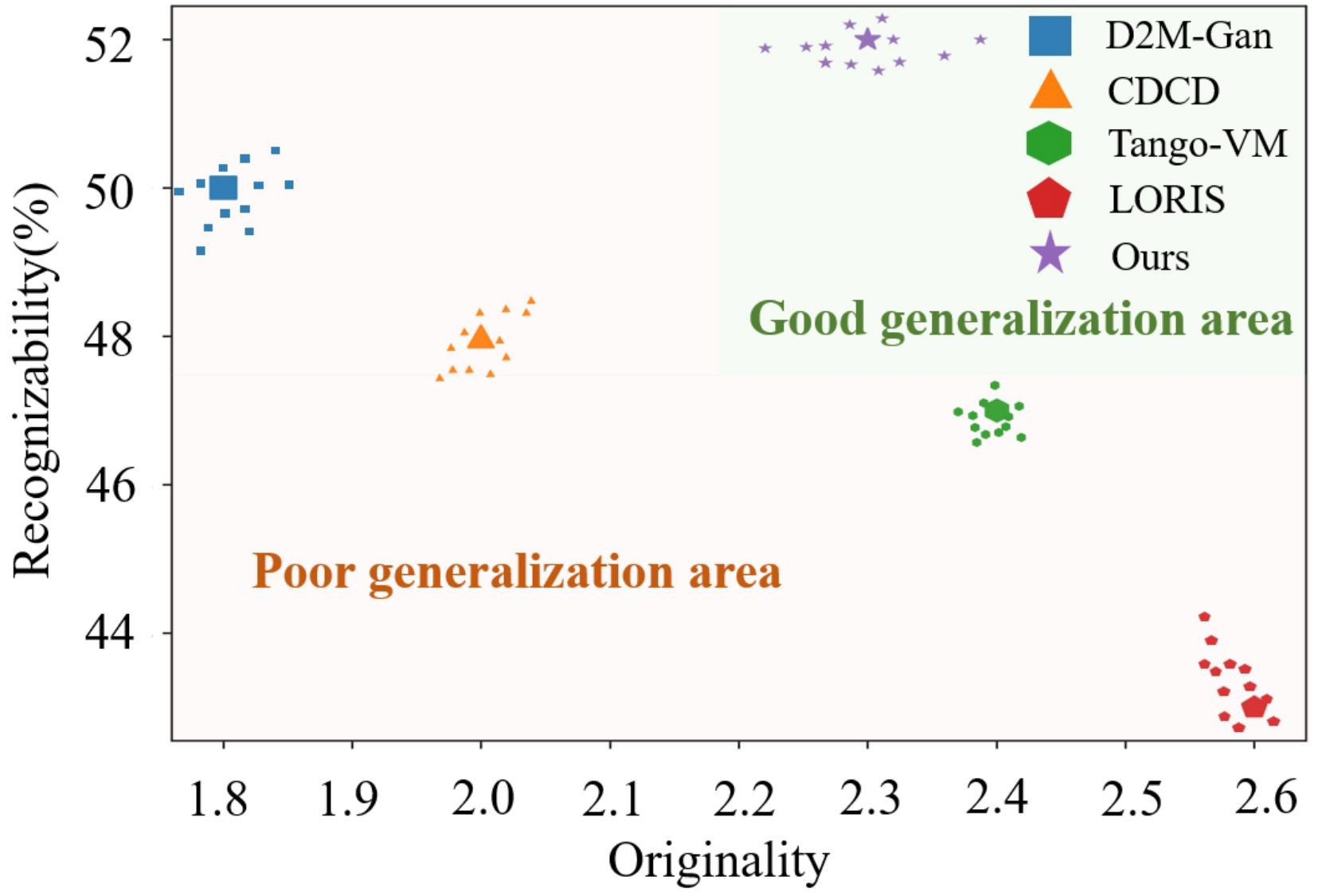}
  \caption{Originality vs. Recognizability Comparison. The big data point represents a single model, encapsulating the average scores of originality and recognizability across all categories,
while the surrounding smaller data points illustrate the scores within each category for that model.}
  \label{fig:R-O}
\end{wrapfigure}\textit{Originality} within film scoring is a critical metric, necessitating the creation of compositions that exhibit distinctiveness when compared with prior background music. This principle underscores the importance of innovation in the composition process, ensuring that each score contributes a unique auditory experience that enhances the narrative and emotional depth of the film. The \textit{recognizability} of a film score refers to its capacity to be discerned by audiences as embodying a specific musical style, thereby facilitating the identification of the score's stylistic lineage or intent. This attribute is pivotal in evaluating the effectiveness of a score in invoking intended emotional responses or thematic associations.
To compute the \textit{Originality}, we use the following equation,
$ \sigma_{\rho} = \sum_{j=1}^{n_c} \sqrt{\frac{1}{N-1} \sum_{i=1}^{N} \left( f(m_{i}^{j}) - \frac{1}{N} \sum_{i=1}^{N} f(m_{i}^{j}) \right)^{2}} 
$,
where ${n_c}$ denotes class number, $f(m_{i}^{j})$ denotes extract feature for mel-spectrogram $m_{i}^{j}$ from pre-trained model.
To evaluate \textit{recognizability}, we use a one-shot classification model designed for musical styles, measuring the classification accuracy of each generated sample. 
The average accuracy across the test set is used as the \textit{recognizability} metric.
Fig.~\ref{fig:R-O} presents a comparison between our approach and these baselines in terms of originality and recognizability. 
In comparison, D2M-Gan~\cite{zhu2022quantized} and CDCD \cite{zhu2022discrete} learned to make identical music of the specific style (i.e., low originality but high accuracy).
Despite their proficiency in accurately learning the distribution of the Ground Truth, which results in high recognizability, they fall short in generating new music compositions.
This suggests that both diffusion models and GANs are unable to overcome the limitations imposed by the Codebook.
The waveform-based LORIS~\cite{Yu2023Long}, by learning from the most original representation of music, is capable of continually creating its own music. However, this leads to the generated music deviating from the distribution of Ground Truth, resulting in lower recognizability.
Our model, along with Tango-VM, effectively balances the relationship between originality and recognizability, ensuring a favorable outcome in both innovation and stylistic expression in music.
Since Tango~\cite{ghosal2023text} has larger parameters and achieves higher spectral resolution, it results in greater creativity compared with our model.

\subsection{Main Results}
\subsubsection{Film Score Generation}

The quantitative experimental results of our film score generation model are shown in Tab.~\ref{tab-eva1}.
Our model outperforms all baselines in terms of rhythm consistency, music quality, and generation stability.
The baselines are designed for dance/sport music generation, focusing on modeling the video pose/motion to the rhythm of music.
It is not suitable for film music with semantic and emotion serveing as bridges to film and score.
To be noted, LORIS~\cite{Yu2023Long} method has the lower BHS and F1 scores compared with other methods. 
The extraction of critical pose features, as a crucial visual rhythm condition for LORIS~\cite{Yu2023Long}, poses challenges within our FilmDB due to the variability and complexity of tasks and scene transitions characteristic of film videos.
Therefore, baselines are designed for dance/sport music generation by aligning video motion to rhythm, fail to cater to film music's unique requirements, where semantic and emotional connections are essential, making them unsuitable for our film scoring context.
Nevertheless, the waveform-based LORIS method outperforms CDCD~\cite{zhu2022discrete} and D2M-Gan~\cite{zhu2022quantized} , as well as the spectrum-based Tango-VM, in terms of overall music quality scores.
This suggests that waveforms synthesized by the generative model better handle musical details, ensuring an improvement in music quality.
From Tab.~\ref{tab-eva1}, we observe that all methods perform well in terms of rhythm consistency on the MV dataset, indicating that films present more challenging rhythm scenarios. 
Nonetheless, our model still achieves significant improvements compared with other methods on both datasets. 
These results validate that our framework can consistently generate high-quality music tracks with accurate rhythm alignment for both films and music videos.
\begin{table}[tb]
\caption{Quantitative evaluation results for the film score generation task on the FilmScoreDB and EmoMV datasets.}
\label{tab-eva1}
\centering
    \scalebox{0.65}{\begin{tabular}{c|c|ccc|ccc|cc|c}
    \toprule
\multicolumn{1}{c|}{\multirow{2}{*}{\textbf{Dataset}}}&\multicolumn{1}{c|}{\multirow{2}{*}{\textbf{Method}}}& \multicolumn{3}{c|}{\textbf{Rhythm Correspondence}} & \multicolumn{3}{c|}{\textbf{Music Quality}} & \multicolumn{2}{c}{\textbf{Generate Stability}} &\multicolumn{1}{|c}{\textbf{Subjective Evaluation}}\\  
\multicolumn{1}{c|}{}    &\multicolumn{1}{c|}{}    & \textbf{BCS↑} & \textbf{BHS↑} & \textbf{F1 scores↑} & \textbf{IS↑} & \textbf{KL↓} & \textbf{FAD↓} & \textbf{CSD↓}       & \textbf{HSD↓}  & \textbf{Mos↑}   \\ \midrule
&D2M-Gan & 60.1 & 61.1  & 62.2   & 1.3  & 5.5   & 14.3   & 24.1  & 25.7 & 2.9  \\
&CDCD   & 57.2   & 60   & 61.1  & 1.6   & 8.6  & 15.6   & 22.4  & 23.8  & 2.7 \\
\multirow{2}{*}{\textbf{FilmScoreDB}}&Tango-VM & 62.1 & 57.3   & 61.4    & 3.7  & 6.7 & 10.8  & 25.4 & 22.5 & 3.3\\
&LORIS & 58.3 & 50.1 & 60.3 & 4.0  & 5.7 & 7.5 & 20.4 & 22.3 & 3.1 \\
&DIFF-Foley & 64.2 & 59.7 & 61.6 & 3.3  & 5.4 & 8.3 & 19.1 & 18.7 & 3.4 \\
\cmidrule{2-11}
&\textbf{HPM} & \textbf{66.7} & \textbf{62.2} & \textbf{64.4}  & \textbf{4.4} & \textbf{5.3} & \textbf{7.1}  & \textbf{18}  & \textbf{16.2} & \textbf{3.9} \\ \midrule
&D2M-Gan       & 62.4  & 63.2 & 65.4  & 1.8   & 5.2  & 12.1  & 22.5  & 23.2  & 3,3\\
&CDCD          & 59.3  & 62.6 & 63.2  & 1.9   & 7.3  & 13.7  & 21.3  & 20.9 & 3.1 \\
\multirow{2}{*}{\textbf{EmoMV}}&Tango-VM         & 64.5  & 59.4 & 63.7  & 4.3   & 5.9  & 8.7  & 23.5  & 20.3  & 3.6\\
&LORIS        & 59.5  & 52.3 & 63.8  & 4.5   & 5.4  & 7.5  & 19.3  & 20.6  & 3.0\\ 
&DIFF-Foley & 67.1 & 60.2 & 64.4 & 3.7  & 5.6 & 9.4 & 18.2 & 17.5 & 3.5 \\
\cmidrule{2-11}
&\textbf{HPM} & \textbf{70.1} & \textbf{64.3} & \textbf{66.2} & \textbf{5.2} & \textbf{4.9} & \textbf{7}  & \textbf{16.7}  & \textbf{15.6}  & \textbf{3.8} \\ \bottomrule

 \end{tabular}}
\end{table}

\begin{figure}[tb]
  \centering
  \includegraphics[height=5cm]{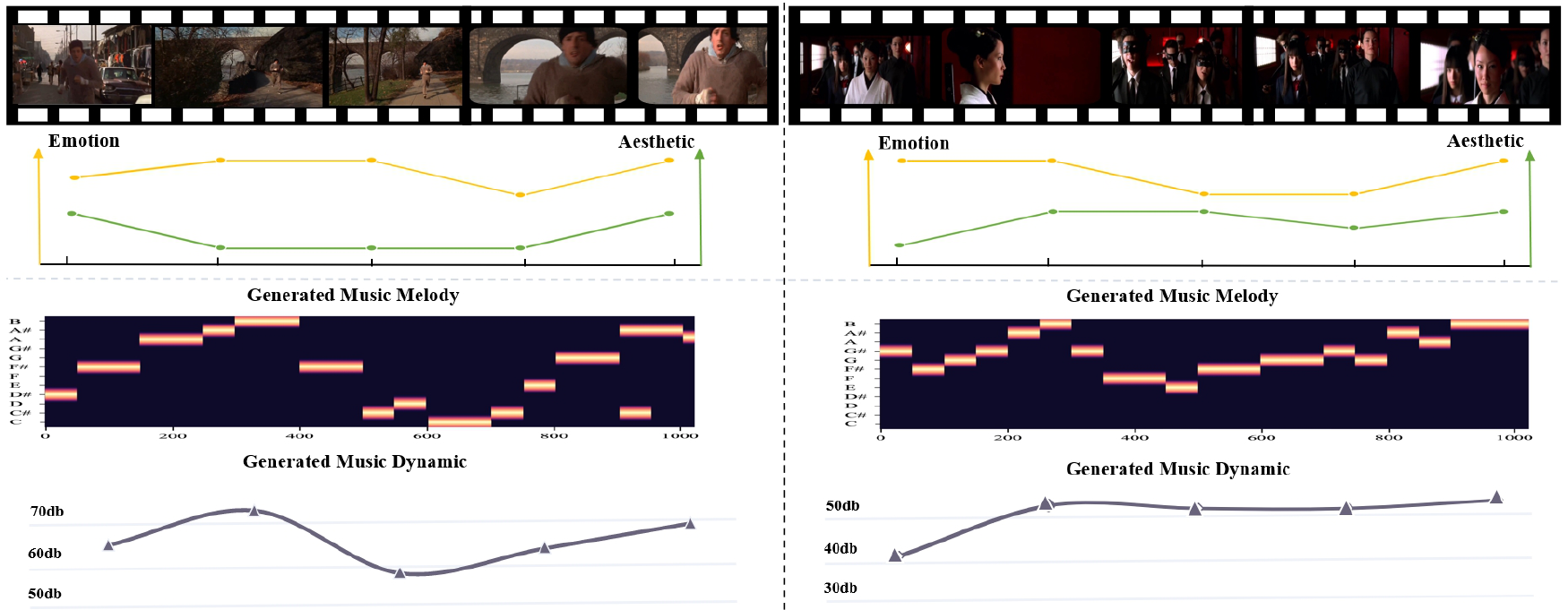}
  \caption{Examples of our Film Score Generation, including video frame (top), emotion \& aesthetic score (middle), and generated melody \& dynamic (bottom).}
  \label{fig:visual2}
\end{figure}

In Fig.~\ref{fig:visual2}, we present examples of generating film scores, showcasing the video frames along with the corresponding emotional and aesthetic scores and the melody and dynamic of the generated music. 
Emotion and aesthetics represent two distinct feature dimensions for characterizing films, with the former reflecting the perceived strength of emotion conveyed by the subjects or the scene, and the latter can refer to the overall composition, balance, and visual appeal of the frame.
The generated dynamics and emotional intensity values are positively correlated.
Emotion and aesthetics collectively determine the melody.

\subsubsection{Composition Style Transfer}
\begin{figure}[tb]
  \centering
  \includegraphics[height=3.2cm]{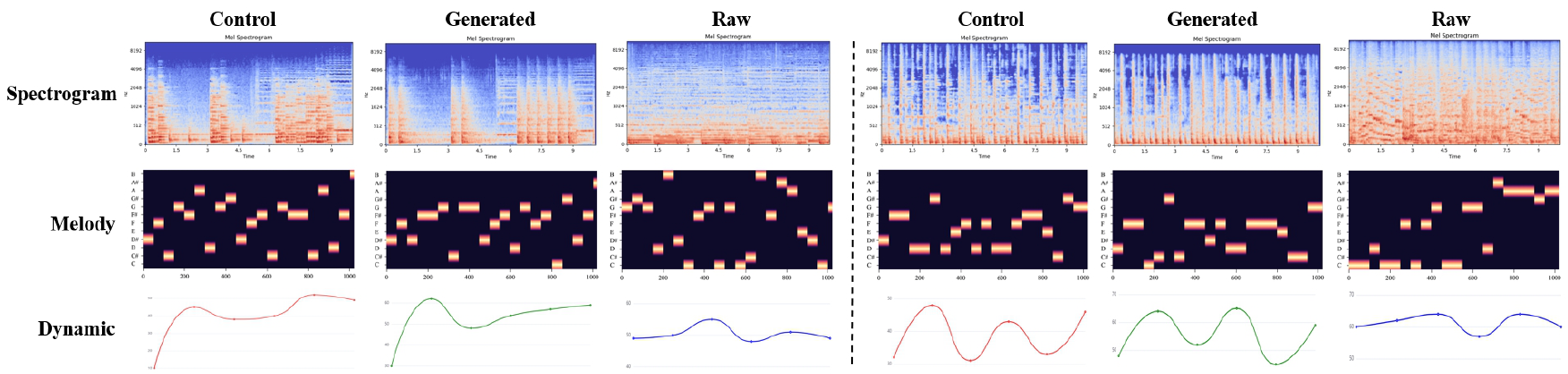}
  \caption{Examples of our composition style transfer, including spectrogram (top), melody (middle), and dynamic (bottom).}
  \label{fig:visual1}
\end{figure}

The quantitative experimental results of our composition style transfer are shown in Tab.~\ref{tab-eva3}.
We establish two style-controlled test sets from the FilmScoreDB and EmoMV datasets. The FilmScoreDB test set categorizes music by film genres (e.g., Plot, Love and Action) and selects control signals of melody and dynamics from different genres to form a diverse control set. 
In contrast, the EmoMV test set, with six emotion categories, uses a similar approach but increases the control samples from 2 to 10 per category to enrich the experimental control set, aiming to assess the model's ability to generate music across varied styles and emotional tones.
Under three different conditions, although there is a significant difference in the quality of generated outputs between the two methods, our control scores are better than DITTO's method.
Additionally, HPM more accurately responds to melody and dynamic controls. 
This suggests that the optimization-based approach during inference is better suited for accurately controlling melody and dynamics in complex music compared with DITTO's method.
In Fig.~\ref{fig:visual1}, we display examples of outputs from models under the control of styles (melody and dynamics) extracted from target music. 
Comparing the mel spectrograms, melody, and dynamic structures of original, generated, and target music reveals that while there may not be complete overlap, the comparison sufficiently shows our model's ability to generate new music in a distinct style from the Ground Truth under specified controls.
\begin{table}[tp]
\caption{Quantitative evaluation results for the composition style transfer task on FilmDB and EmoMV datasets.}
\label{tab-eva3}
\centering
\scalebox{0.7}{\begin{tabular}{c|c|c|c|cc|ccc}
\toprule
\multirow{2}{*}{\textbf{Dataset}} &\multirow{2}{*}{\textbf{Control}} & \multirow{2}{*}{\textbf{Model}} & \multirow{2}{*}{\textbf{Melody acc (\%)}} & \multicolumn{2}{c|}{\textbf{Dynamics corr(r, in \%)}} & \multicolumn{3}{c}{\textbf{Music Quality}}  \\
& & & & \textbf{Micro} & \textbf{Macro}& \textbf{IS↑} & \textbf{KL↓} & \textbf{FAD↓} \\ \midrule
&Melody & DITTO  & 51.3   & 1.7  & 3.2   & 4.8   & 4.2   & 6.1  \\
&only   & \textbf{Ours}   & \textbf{57.6}   & \textbf{2.8}  & \textbf{3.9}   & \textbf{5.8}   & \textbf{5.4}   & \textbf{7.4}  \\ \cmidrule{2-9}
\multirow{2}{*}{\textbf{FilmScoreDB}}&Dynamics & DITTO & 4.5  & 54.3  & 79.3   & 5.1   & 4.5   & 5.9  \\
&only  & \textbf{Ours} & \textbf{8.4}  & \textbf{60.3} &  \textbf{84.7}  & \textbf{5.4}   & \textbf{5.3}   & \textbf{7.2}   \\ \cmidrule{2-9}
&Melody & DITTO & 52.3 & 61.7 & 82.4 & 5.4 & 4.8 & 6.7 \\ 
&Dynamics  & \textbf{Ours} & \textbf{57.7}  &  \textbf{64.5} & \textbf{87.2} & \textbf{6.1} & \textbf{5.6} &\textbf{7.9} \\ \cmidrule{1-9}
&Melody & DITTO  & 52.5   & 2.4  & \textbf{5.2}   & 5.3   & 4.7   & 6.8  \\
&only   & \textbf{Ours} & \textbf{58.2}& \textbf{3.6}& 5.1 & \textbf{6.4}& \textbf{5.2} & \textbf{7.9}  \\ \cmidrule{2-9}
\multirow{2}{*}{\textbf{EmoMV}}&Dynamics & DITTO & 4.8  & 55.3  & 79.9   & 5.7   & 5.3   & 6.4  \\
&only  & \textbf{Ours} & \textbf{9.6}  & \textbf{63.4} &  \textbf{86.8}  & \textbf{6.2}   & \textbf{5.4}   & \textbf{7.7}   \\ \cmidrule{2-9}
&Melody & DITTO & 53.4 & 62.5 & 83.4 & 6 & 5.4 & 7.2 \\ 
&Dynamics  & \textbf{Ours} & \textbf{58.6}  &  \textbf{66.3} & \textbf{89.1} & \textbf{6.7} & \textbf{5.6} &\textbf{8.3} \\ \bottomrule
\end{tabular}}
\end{table}
\subsubsection{Low-rank Adaptation for Accelerated Training} To prove the efficiency of
LORA, we evaluate two model variants: HPM with/without LORA in Tab.~\ref{tab:sub1}, noting differences in trainable parameters and training durations. 
Remarkably, HPM with LORA reduces parameters from 87 million to 20 million and cuts training time from 48 to 12 hours without compromising performance.  
HPM without LORA marginally outperformed in melody accuracy, while the HPM with LORA has slightly better dynamics correlation scores.
This indicates LORA's efficacy in streamlining the model and quickening training while maintaining or slightly improving dynamics assessment.

\begin{table}[t]
\centering
\caption{The results of the ablation study of Film Encoder and the comparative experiment of the accelerated with LORA on the FilmScoreDB dataset.}
\begin{subtable}{.5\linewidth}
\centering
\caption{The impact of LoRA.}
\scalebox{0.85}{
\resizebox{\textwidth}{15.2mm}{
\begin{tabular}{c|ccc}
\toprule
\textbf{Method} & \multicolumn{2}{c}{\textbf{Trainable Parameters}} & \textbf{Train Time} \\ \midrule
HPM w/o LORA & \multicolumn{2}{c}{87M}  & 48 hours   \\
HPM w/ LORA  & \multicolumn{2}{c}{\textbf{20M}}  & \textbf{12 hours}  \\ \bottomrule
\multirow{2}{*}{\textbf{Method}}  & \multicolumn{1}{c}{\multirow{2}{*}{\textbf{Melody acc(\%)}}} & \multicolumn{2}{c}{\textbf{Dynamics corr(r,in \%)}} \\
& \multicolumn{1}{c}{}  & \textbf{Micro}  & \textbf{Macro}  \\ \midrule
\multicolumn{1}{l|}{HPM w/o LORA} & \multicolumn{1}{c}{\textbf{57.7}}   & 64.5     & 87.2   \\
HPM w/ LORA     & \multicolumn{1}{c}{57.3}        & \textbf{64.7}   & \textbf{87.6}   \\ \bottomrule
\end{tabular}}}

\label{tab:sub2}
\end{subtable}%
\begin{subtable}{.5\linewidth}
\centering
\caption{Different Film Encoder conditions.}
\scalebox{0.75}{
\setlength{\tabcolsep}{2.2mm}{
\resizebox{\textwidth}{17mm}{
\begin{tabular}{c|ccc|cc}

    \toprule
\multicolumn{1}{c|}{\multirow{2}{*}{\textbf{Method}}} & \multicolumn{3}{c|}{\textbf{Music Quality}} & \multicolumn{2}{c}{\textbf{Generate Stability}} \\ 
\multicolumn{1}{c|}{}   &  \textbf{IS↑} & \textbf{KL↓} & \textbf{FAD↓} & \textbf{CSD↓}  & \textbf{HSD↓}  \\ \midrule
HPM w/ \textbf{S}, \textbf{A}  & 4.6 & 5.5 & 7.3 & 18.1 & 16.1 \\
HPM w/ \textbf{A}, \textbf{E} & 4.5 & 5.4 & 7.2 & 17.9 & 16.0 \\
HPM w/ \textbf{E}, \textbf{S}  & 4.5 & 5.4 & 7.2 & 18.0 & 16.1 \\
HPM w/ \textbf{S}  & 4.5 & 5.4 & 7.2 & 17.8 & 16.0 \\
HPM w/ \textbf{E}  & 4.6 & 5.5 & 7.3 & 18.0 & 16.1 \\
HPM w/ \textbf{A}  & 4.7 & 5.6 & 7.4 & 18.3 & 16.2 \\ \midrule
\textbf{HPM}  & \textbf{4.4} & \textbf{5.3} & \textbf{7.1}  & \textbf{18}  & \textbf{16.2}  \\ \bottomrule
\end{tabular}}}
}
\label{tab:sub1}
\end{subtable}%

\label{tab:combined}
\end{table}

\subsection{Ablation Study}
We conduct an ablation study to test the influence of film encoder and film score ControlNet on the FilmScoreDB dataset.
For the film encoder, we test different combinations of visual conditions (\textbf{S}emantic, \textbf{A}esthetic, \textbf{E}motion) on the performance of the HPM model in Tab.~\ref{tab:sub2}.
Each row represents a variant of the model with specific combinations of visual conditions.
We observe that incorporating different combinations of visual conditions leads to variations in performance across all evaluation metrics compared with the complete model (HPM).
Models with combined visual conditions generally exhibit higher performance compared with models with individual visual condition.
The performance of the HPM model with different combinations of visual conditions remains relatively consistent across various metrics, indicating that each combination contributes to enhancing music generation tasks.
Overall, these findings highlight the importance of considering multiple visual conditions simultaneously to achieve optimal performance in music generation tasks using the HPM model.

\section{Discussion}
In this paper, we investigate the task of automatic film scoring through the introduction of a comprehensive 90.35-hour film-music dataset, alongside the establishment of a baseline model (HPM) and an evaluative metric focus on \textit{Originality vs. Recognizability}. We acknowledge certain constraints and delineate avenues for future enhancement: A primary constraint of our current framework is rigidity in producing music of fixed durations, limiting its adaptability to diverse emotional expressions, narrative forms, or situational demands. We intend to adjust sampling rates and density to vary output lengths while preserving the music's integrity.
\\

\bibliographystyle{splncs04}
\bibliography{neurips_2024}
\newpage

\appendix
\begin{table}[tb]
\centering
\caption{Overview of the video2music method.}
\label{tab-overview}
    \scalebox{0.65}{\begin{tabular}{cccccc}
    \toprule
    \textbf{Model}& \textbf{Music Representation} & \textbf{Architecture} & \multicolumn{2}{c}{\textbf{Mertics}}& \textbf{Dataset}\\ \midrule
    Foley Music~\cite{gan2020foley} & MIDI& Transformer& \multicolumn{2}{c}{NDB}& URMP,Atin Piano,MUSIC  \\
    CMT~\cite{di2021video} & MIDI & Transformer & \multicolumn{2}{c}{PCHE,GPS,ST} & LPD  \\
    D2M-Gan~\cite{zhu2022quantized} & Codebook  & Gan & \multicolumn{2}{c}{BCS,BHS,Genre Accurary,MOS} & AIST++,TikTok \\
    CDCD~\cite{zhu2022discrete}  & Codebook  & Diffusion & \multicolumn{2}{c}{BCS,BHS,Genre Accurary,MOS} & AIST++,TikTok \\
    V-Musprod~\cite{zhuo2023video} & MIDI  & Transformer  & \multicolumn{2}{c}{VMCP,SC,PE,PCE,EBR,IOI} & SymMV \\ 
    LORIS~\cite{Yu2023Long} & Waveform & Diffusion & \multicolumn{2}{c}{MOS,BCS,BHS,F1 Scores} & FisV,AIST++,FS1000,Finegym \\
    Video2Music~\cite{kang2023video2music} & MIDI & Transformer  & \multicolumn{2}{c}{ Hits@k scores,EML} & MuVi-sync \\ 
    DIFF-Foley~\cite{luo2024diff} & Mel-Spectrogram & Diffusion  & \multicolumn{2}{c}{IS,FID,MKL,Align Accs} & VGGSound,AudioSet \\ \midrule
    \textbf{Ours} & \textbf{Mel-Spectrogram} & \textbf{ControlNet} & \multicolumn{2}{c}{\textbf{MOS,BCS,BHS,F1 Scores,FAD,KL,OR}} & \textbf{FilmScoreDB(ours),EmoMV} \\ \bottomrule 
  \end{tabular}}
\end{table}
\section{Related Work}
\subsection{Video2Music Generation}
As illustrated in Tab.~\ref{tab-overview}, there has been some progress in the video-to-music generation task. Some works~\cite{gan2020foley,di2021video,zhuo2023video,kang2023video2music} are based on the transformer architecture to generate the MIDI file for music generation autoregressively. However, the limitation lies in the simplicity of the music recorded in the MIDI file, which prevents the model from generating complex music. Some pipelines~\cite{zhu2022quantized,zhu2022discrete} use the Codebook as music representation but are limited by their diversity due to the Codebook. LORIS~\cite{Yu2023Long} is a long-term music generation framework that utilizes a Latent Diffusion Model~\cite{rombach2022high}, mainly focused on figure skating and dance video soundtracks, and lacks other application scenarios. However, DIFF-Foley~\cite{luo2024diff} utilizes the temporal and semantic alignment features learned through contrastive audio-visual pre-training (CAVP) as conditions and then uses LDM to generate the Mel-spectrogram. It effectively demonstrates the ability of the diffusion model and Mel-spectrogram representation in the video-to-music task. Unlike prior works, our framework is capable of automatically generating complex film scores at affordable costs.

\subsection{Controllable Generation}
Although conditional guided generation has achieved impressive results in diffusion, text or video alone cannot finely control the generated content, so controllable generation is gradually emerging. \textbf{ControlNet}~\cite{zhang2023adding} uses lightweight adapters to incorporate conditional inputs (Canny edges, Hough lines, user scribbles, etc.) into latent diffusion models, which can better control the image generation process and generate more specific and compliant images. 
\textbf{Uni-ControlNet}~\cite{zhao2024uni} can accept multiple pixel-level controls through a single adapter branch without specifying all controls simultaneously, while ControlNet requires a separate adapter branch for each control. Meanwhile, \textbf{Inference-time Guidance-based~\cite{hertz2022prompt,bar2023multidiffusion}/Optimization-based~\cite{DBLP:conf/iccv/WallaceGEN23,novack2024ditto} Control} allows for real-time adjustments in the direction of the generative process, offering users greater control over the output.

\begin{figure}
    \centering
    \begin{minipage}{0.48\textwidth}
        \centering
        \includegraphics[width=\linewidth]{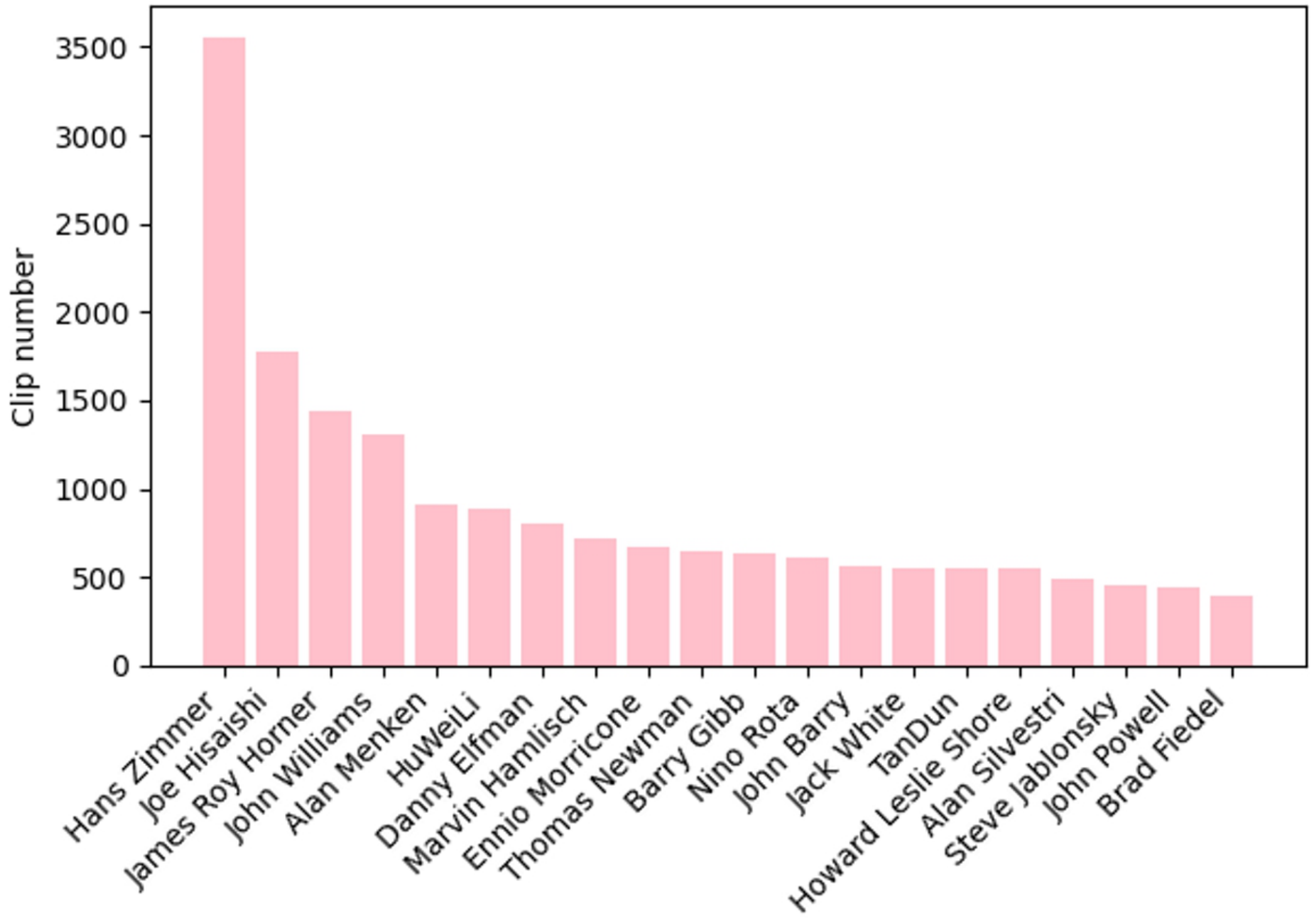}
        \caption{Top 20 composers data distribution}
        \label{fig:composer}
    \end{minipage}
    \hfill
    \begin{minipage}{0.47\textwidth}
        \centering
        \includegraphics[width=\linewidth]{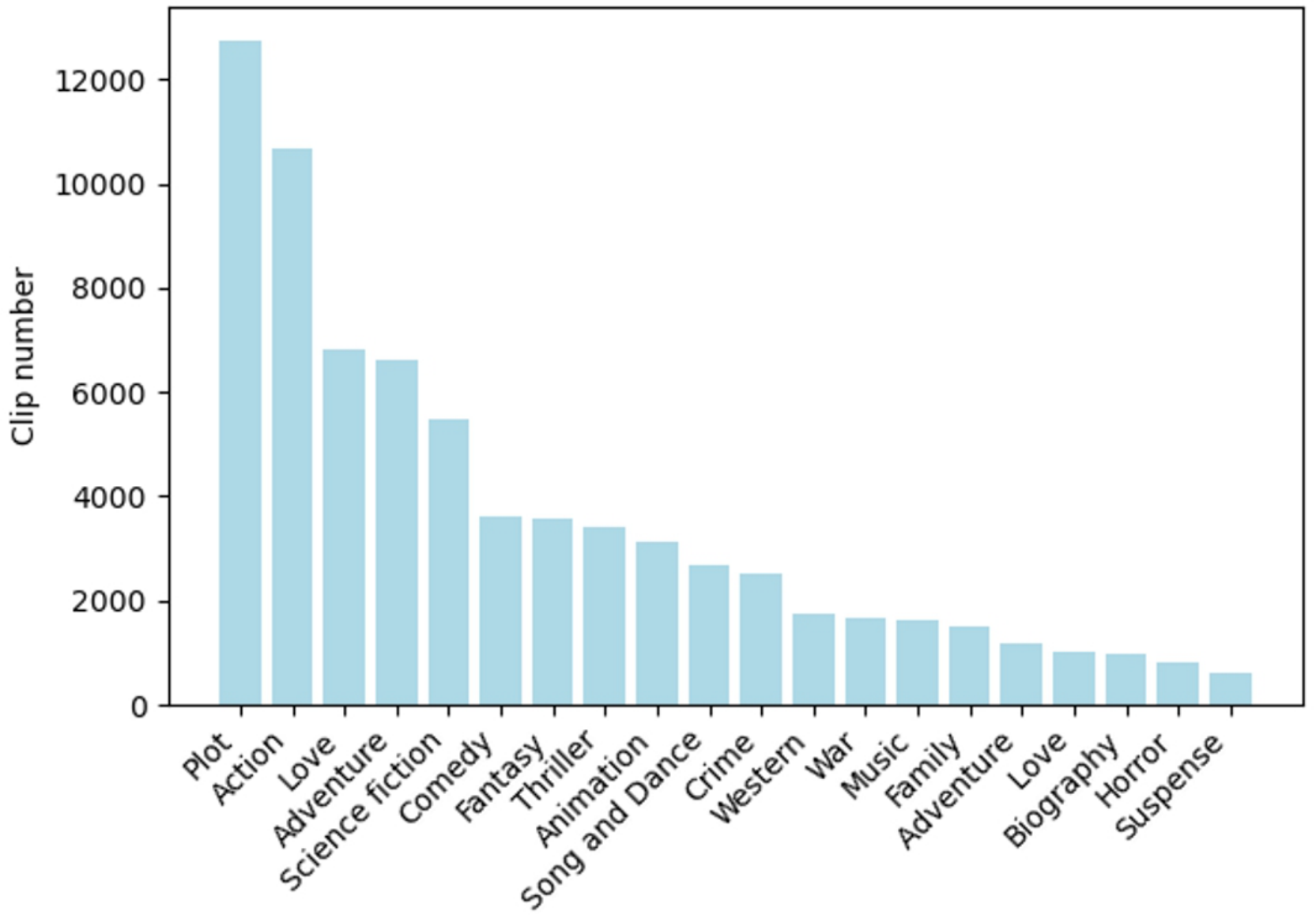}
        \caption{Top 20 movie styles data distribution}
        \label{fig:style_film}
    \end{minipage}
\end{figure}
\section{Dataset}
\subsection{FilmScoreDB}
We annotate 280 films and and obtain 4887 raw film segments. 
The duration of these segments ranges from 5 seconds to 18 minutes. 
For our film score generation, we discard segments less than 10 seconds in duration and split those greater than 10 seconds into 10-second intervals.
Our dataset covers 134 renowned composers, including Hans Zimmer, John Williams, and Danny Elfman, among others.
These composers are universally recognized as top film composers and their works frequently receive nominations for Oscars and Grammys, which ensures the high quality of our film scoring dataset.
In Fig.~\ref{fig:composer}, we present the distribution of sample data for the top twenty composers.
The distribution demonstrates a long-tail effect, encompassing composers who have fewer but highly classic works.
This diversity assists the model in learning film score across various styles and techniques.
In Fig.~\ref{fig:style_film}, we depict the distribution of sample data for the top twenty film genres.
Our dataset encompasses a wide range of film genres, including plot, action, love, and more.
Typically, each film has multiple film genre annotations.
Although dominant genres have a higher proportion, the long-tail effect ensures the inclusion of numerous niche film genres, facilitating the capture of more subtle differences and challenging scenarios. Overall, our dataset is sizable, with substantial data for top composers and film genres, providing ample support for model training.

\begin{figure}[]
  \centering
  \includegraphics[height=7cm]{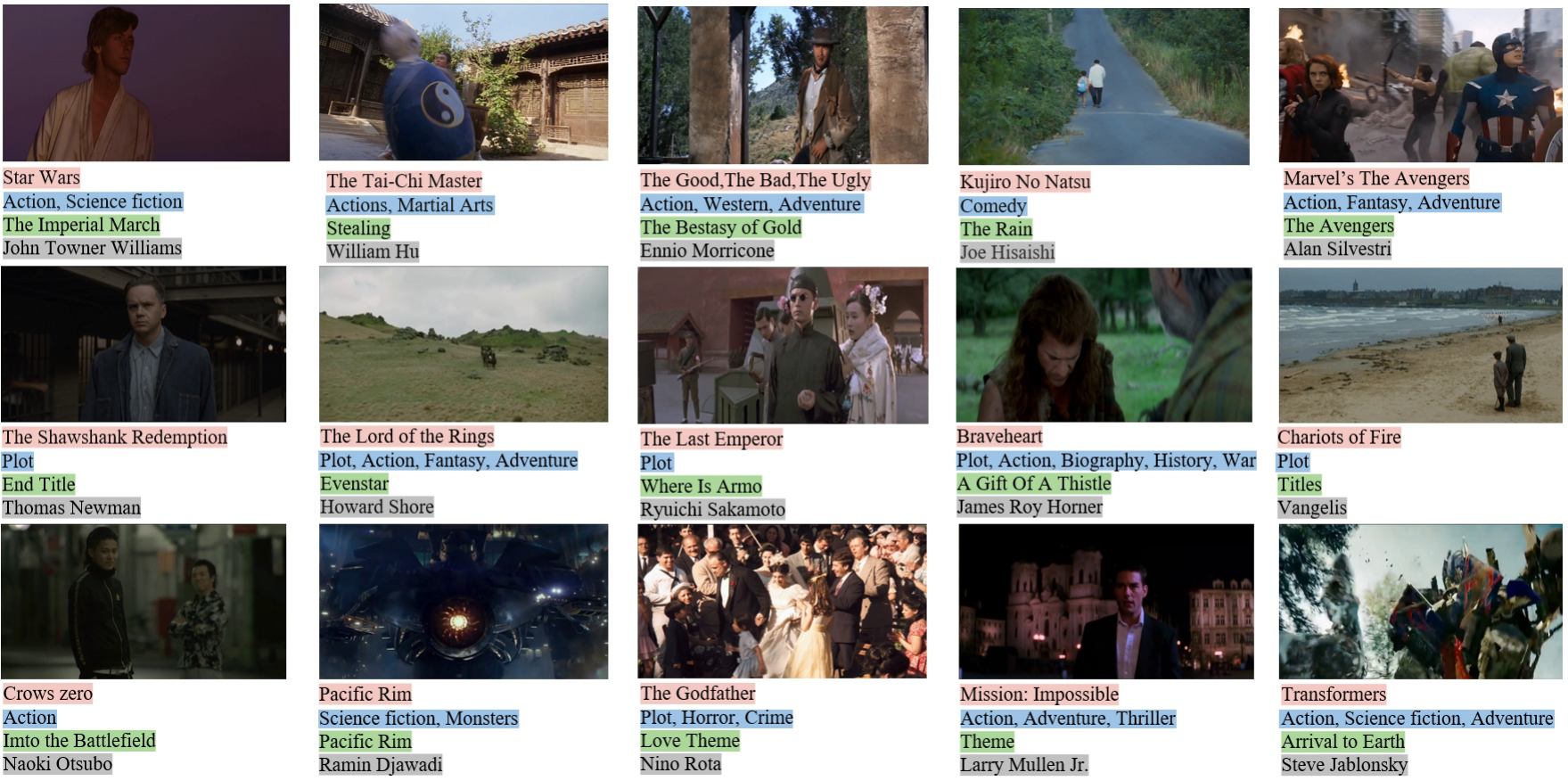}
  \caption{\textbf{Illustration of samples in FilmScoreDB.} The information displayed for each sample includes the movie title, music score title, and composer.}
  \label{fig:FilmScoreDB}
\end{figure}

In Fig.~\ref{fig:FilmScoreDB}, we present a visualization of our FilmScoreDB dataset, which comprises video-music pairs.
Each sample contains not only the movie title, audio, video, music score title, and composer, but also the corresponding movie genre label.
Our dataset encompasses a diverse range of movie categories. At the same time, we also eliminate all vocal content from the audio data using Demucs\footnote{https://github.com/facebookresearch/demucs} to enhance the quality of the model-generated outputs. 
Demucs is a highly effective Music Source Separation model that separates the input music into accompaniment and vocal files.
Here, we select the accompaniment as our final audio data.
We do not own any copyrights to the films and music referenced in the dataset.

\subsection{HIMV and EmoMV}
The HIMV dataset\cite{hong2017content} is a large-scale music video dataset that provides approximately 100k YouTube links. We utilize these links to download around 40k music videos with Yt-dlp\footnote{https://github.com/yt-dlp/yt-dlp}. Subsequently, we select two 10-second segments from the middle of each video to construct an approximate 80k video-music dataset. These data are preprocessed using Demucs and used to train our VM-Net.

The EmoMV dataset consists of data from the MVED dataset\cite{pandeya2021deep}, the Music Mood Dataset\cite{ccano2017music}, and some original music videos from movies collected on YouTube. In total, the dataset includes 5,370 30s data points and 616 10s data points. Subsequently, we clip the 30s data into three segments, resulting in a final set of 16,726 video-music pairs for experimentation.

\section{Film Encoder}
\subsection{Semantic Feature}
CLIP~\cite{radford2021learning} is a pre-trained model for language-image tasks, consisting of a text encoder $f_{text}(\cdot)$ and an image encoder $f_{image}(\cdot)$. It is trained on large-scale image-text datasets using contrastive learning to enhance the similarity between images and text. As a result, CLIP\footnote{https://v-iashin.github.io/video\_features/models/clip} can generate semantic priors not only for images but also for videos, which can be utilized to facilitate music generation.

To generate video semantic priors, we first extract 512-dimensional features from video frames using $f_{image}(\cdot)$ at a frame rate of 10 frames per second. Then, we average the feature sequence along the time dimension to obtain the visual semantic feature $c_s \in R^{1\times512}$ of the entire video. The $c_s$ contains different semantic representations in the video, such as scenes, characters, and objects, which can guide the generation of music.

\subsection{Aesthetic Feature}
We employ the TAVAR\footnote{https://github.com/yipoh/TAVAR} model~\cite{li2023theme} for aesthetic feature extraction. It comprises three key components: The Visual Attribute Analysis Network (VAAN), The Theme Understanding Network (TUN), Bilevel Aesthetic Reasoning.

1)Visual Attribute Analysis Network (VAAN), which is used to learn visual attributes for aesthetic perception. It uses a ResNet50~\cite{he2016deep} architecture with the fully connected layers removed, allowing feature extraction to be shared across branches. Then, we utilize six Multilayer Perceptrons (MLPs) to further map the shared features to six visual attributes. Specifically, we extract $10$ frames from our movie clip, denoted as $a_i(i=1,2,...,10)$, at a frame rate of 1 frame per second. For each input image $a_i$, the hidden feature $h_a^i$ is obtained from the shared feature extraction network $F_a(\cdot)$ as:
\begin{equation}
    h_a^i = F_a(a_i).
\end{equation}
Then, six MLPs are used to construct six attribute branches, further mapping the hidden features $h_a^i$ of each image to visual attributes $\hat{a}^i$, which are defined as follows:
\begin{equation}
    \hat{a}^i = MLP_m(h_a^i),
\end{equation}
where $m=1,2,\ldots,6$  denotes $6$  attribute branch MLPs, and $\hat{a}^i = \left\{\hat{a}_1^i,\hat{a}_2^i,\ldots,\hat{a}_6^i\right\}$ denotes six predicted visual attributes.
We take the average of 10 frames to obtain 6 distinct attributes (the mean of different video frames) as the visual attributes $\hat{a}_j$ for the entire video. These visual attributes are then used as input for the module of Bilevel Aesthetic Reasoning.

2)Theme Understanding Network (TUN), which utilizes a ResNet-50 backbone~\cite{he2016deep} to initially predict the theme category of an image. Then, a Multi-Layer Perceptron (MLP) with PReLU activation function maps the input image to the predicted theme category, where the last fully connected layer produces six outputs representing six theme categories. Finally, a softmax non-linearity operation is performed to generate the predicted theme probabilities. Since this module not only predicts the theme category of an image but also generates theme features, we feed the 10 frames of a movie video into this module to extract theme features for each frame, which are defined as:
\begin{equation}
    \hat{b}^i = MLP(F_t(a_i)).
\end{equation}
Then, we average the theme features of these 10 frames to obtain the theme feature $\hat{b}$ for the movie video. This feature is also inputted into the module of Bilevel Aesthetic Reasoning.
3)Bilevel Aesthetic Reasoning. Firstly, considering the relationship between image themes and visual attributes, the theme features serve as central nodes, and all attribute nodes are connected to the central theme node. Finally, node features are obtained through a graph convolutional network(GCN). Additionally, based on the relationship between theme-aware visual attributes and general aesthetics, general aesthetic features are treated as the central node, and all theme-aware visual attribute nodes are connected to the central node. By utilizing the attribute aesthetics graph, updated node features that integrate theme-aware visual attributes and aesthetic features are obtained. Here, aesthetic features are extracted using a Swin Transformer~\cite{liu2021swin}.

Lastly, we append an FC layer to map the updated node features to an overall aesthetic quality score. The overall aesthetic quality score is then embedded to obtain aesthetic features $c_a \in R^{1\times512}$, which guide music generation.

\subsection{Emotion Feature}
Here, we employ a pre-trained Weakly Supervised Video Emotion Detection and Prediction(WECL\footnote{https://github.com/nku-zhichengzhang/CTEN}) model~\cite{zhang2023weakly}, to predict emotional information within the videos.
We embed one-hot categorical labels E into the Emotion feature $c_e \in R^{1\times512}$ via linear projection:
\begin{equation}
    c_e = Embed(E).
\end{equation}

\subsection{Feature Fusion}
For the semantic feature $c_s$, aesthetic feature $c_a$, and emotional feature $c_e$, if we use the typical feature vector concatenation, we will face the curse of dimensionality and a lack of interaction between features. Therefore, we have employed a simplified feature fusion block called Lightweight Attention Feature Fusion (LAFF)~\cite{hu2022lightweight}. In this approach, we combine the semantic, emotional, and aesthetic features of the video in a specific LAFF block using a convex combination, where the fusion weights are learned to optimize the generation of movie music.

\section{Objective Metric}
\subsection{BCS, BHS, F1 scores, CSD and HSD}
Here, we use improved versions of the Bar Coverage Score (BCS) and Bar Hit Score (BHS) for rhythm relevance evaluation. BCS and BHS are typically evaluated by computing the aligned rhythm beats between synthesized and real music, but due to these metrics being only applicable to short-length music (2~6s), two main issues arise when evaluating long music: 1) The second-wise rhythm detection algorithm would lead to extremely sparse vectors for any long music sequences, thus consistently low BCS and BHS values fail to reflect the actual performance. 2) If the generated music involves more rhythm beats than the ground truth, BHS can easily exceed 1, which would make the metric values appear perfect while the per-sample performance is far from satisfactory. We use two modified BCS and BHS similar to LORIS~\cite{Yu2023Long}: First, adjust the parameters of the audio onset detection algorithm to avoid sparse rhythm vectors. Second, compute BCS by dividing the aligned beats by the total beats of the generated music $(B_a/B_g)$, where BCS and BHS serve as recall and precision respectively. Additionally, we use the F1 scores of BCS and BHS as a comprehensive evaluation, and CSD and HSD (the standard deviations of BCS and BHS) to evaluate generation stability.
\subsection{Melody acc and Dynamics corr}
Melody Accuracy assesses the match between frame-wise pitch classes ($C, C\#$,
$\ldots , B$; totaling 12) of input melody controls and those extracted from the generated output. Dynamics Correlation measures the Pearson correlation between frame-wise input dynamic values and those computed from the generation. Two types of correlations are computed: micro and macro. Micro calculates r's for each generation separately, while macro collects input/generation dynamic values across all generations and computes a single r. Micro correlation examines whether relative dynamic control values are respected within a generation, while macro correlation assesses the same attribute across multiple generations.
\section{Additional Examples}
\begin{figure}[htbp]
  \centering
  
  \begin{minipage}[t]{1\linewidth}  
      \centering
      \includegraphics[height=4cm]{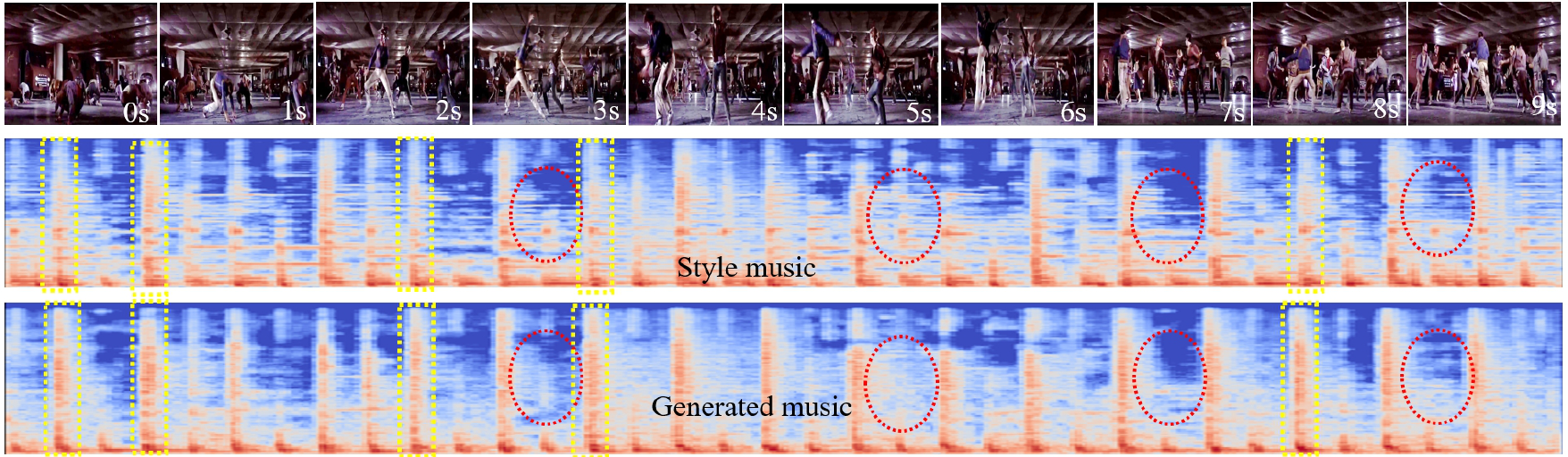}
      \subcaption{}
  \end{minipage}
  \begin{minipage}[t]{1\linewidth}
      \centering
      \includegraphics[height=3.8cm]{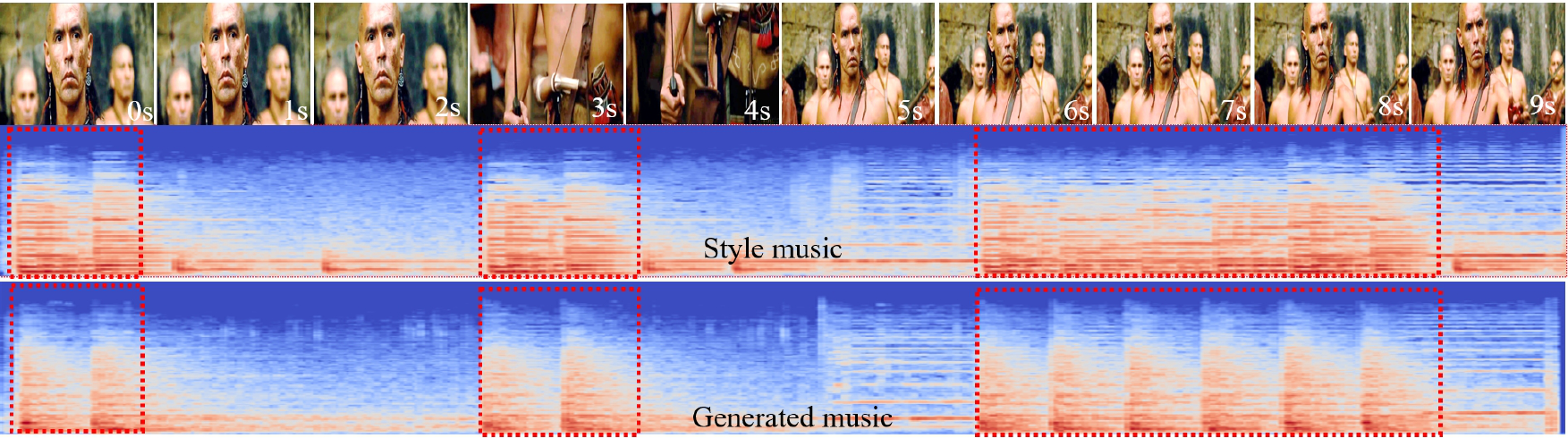}
      \subcaption{}
  \end{minipage}
  \begin{minipage}[t]{1\linewidth}
      \centering
      \includegraphics[height=3.8cm]{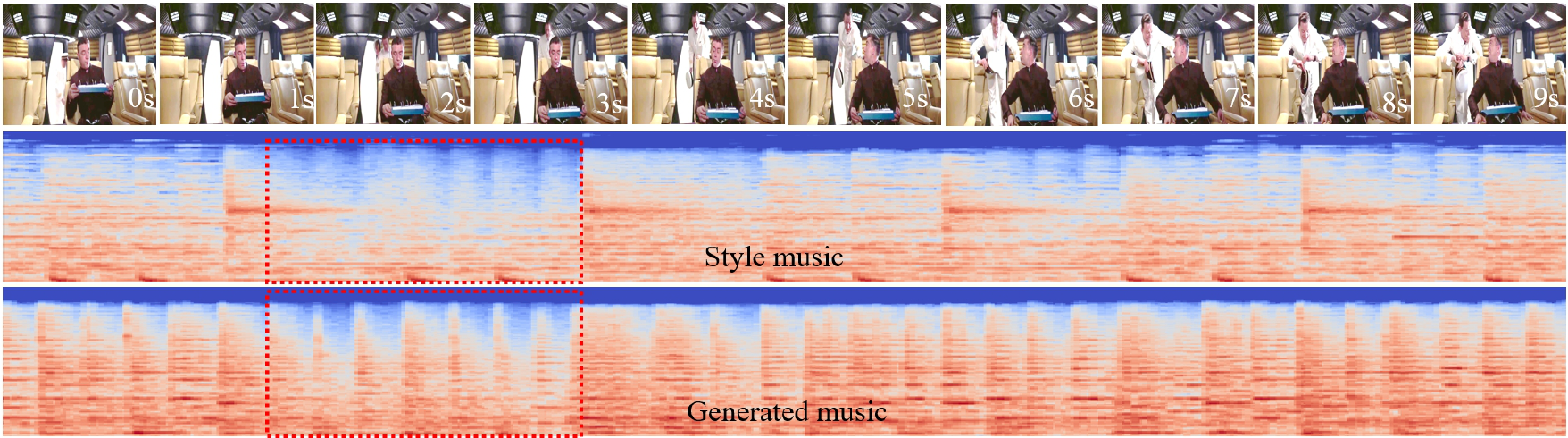}
      \subcaption{}
  \end{minipage}
  \caption{More Composition Style Transfer Results on FilmScoreDB. We visualize the spectrograms of Style Music and Generated Music.}
  \label{fig:demo}
\end{figure}

In Fig.~\ref{fig:demo}, we show three additional examples of composition style transfer with HPM. In these three examples, our model can highlight the melody of style music (red rectangle) by learning the main melody of style music while enhancing music segments that match the video dance beat (circled in yellow) and eliminating noise between different notes (circled in red). This makes the generated music more compact and compatible with video content.
\end{document}